\documentclass[aps,prl,reprint,showpacs,groupedaddress]{revtex4-1}
\usepackage{color}
\usepackage{graphicx}
\usepackage{bm}
\usepackage{amsmath}
\usepackage{amssymb}

\newcommand{\eqn}[1]    {(\ref{#1})}
\newcommand{\fig}[1]	{Fig.~\ref{#1}}

\begin{document}


\title{Bifurcation of the  Edge-State Width in the Two-Dimensional Topological Insulator}

\author{Hyeonjin Doh}
\email[]{clotho72@yonsei.ac.kr}
\affiliation{Department of Physics and Center for Computational Studies of 
Advanced Electronic Material Properties, Yonsei University, Seoul 120-749, Korea}

\author{Gun Sang Jeon}
\email[]{gsjeon@ewha.ac.kr}
\affiliation{Department of Physics, Ewha Womans University, Seoul 120-750, Korea}


\date{\today}

\begin{abstract}
We examine the properties of edge states in a two-dimensional topological
insulator.
Based on the Kane-Mele model, we derive two coupled equations
for the energy and the effective width of edge states at a given 
momentum in a semi-infinite honeycomb lattice with a zigzag boundary.
It is revealed that, in a one-dimensional Brillouin zone,
the edge states merge into the continuous bands of the bulk states 
through a bifurcation of the edge-state width.
We discuss the implications of the results to the experiments in monolayer
or thin films of topological insulators.
\end{abstract}


\maketitle

Topological insulators (TI) are one of the fascinating
fields which have attracted extensive studies 
in condensed matter physics for the past decade~\cite{Hasan2010,Qi2011}.
This phenomenon can be traced back to quantum Hall effect (QHE) in
two-dimensional (2D) systems under high magnetic fields. 
The QHE is characterized by the presence of gapless edge states with 
a finite gap in the bulk.
This metallic edge channel is revived in TI with 
time-reversal symmetry (TRS) preserved.
This edge state is also known to be topologically protected
as in the QHE.

The study of TI was first initiated theoretically in 2D 
systems~\cite{Kane2005,Bernevig2006} dubbed as
a quantum spin Hall effect,
where spin-orbit coupling(SOC) plays an important
role~\cite{Haldane1988,Zhang2001,Murakami2003,Sinova2004}.
It was subsequently generalized to the TI in 3D systems
with a single Dirac-cone dispersion on the
surface~\cite{Fu2007,Moore2007,Roy2009}.
Finally, the TI state in 2D systems has been confirmed experimentally 
by transport measurements
in HgTe/CdTe quantum well \cite{Konig2007,Konig2008}.
The 3D topological insulators have also been supported experimentally 
by angle-resolved photoemission spectroscopy(ARPES) in
Bi$_x$Sb$_{1-x}$ \cite{Hsieh2008}
, Bi$_2$Se$_3$ \cite{Xia2009}, and Bi$_2$Te$_3$ \cite{Chen2009,Hsieh2009}.
This is consistent with
 the density of states of the metallic surface observed by
scanning tunneling microscopy (STM) for Bi$_{1-x}$Sb$_x$ \cite{Roushan2009},
and Bi$_2$Te$_3$ \cite{Alpichshev2010}.

Although the ARPES and STM measurements support the surface states in TI,
the transport measurements suffer from the residual carriers in the bulk states.
The residual carriers are attributed to the imperfection of the bulk 
crystals such as antisites and vacancies.
There have been many efforts for the reduction of the bulk
carrier density,
which includes chemical doping on Bi$_2$Se$_3$ with Sb~\cite{Analytis2010a}
or on Bi$_2$Te$_3$ with Sn~\cite{Chen2010}.
In spite of some improvements, the reduction of the bulk carrier density is not
sufficient to suppress completely the bulk contribution of the transport 
due to the difficulty of the fine tuning of the doping concentration.
An alternative way to control the bulk carrier density is to reduce the 
sample size or to apply gate voltage. 
Epitaxially grown thin films of Bi$_2$Se$_3$ showed weak antilocalization
effects with large magnetic field, which represents the surface state of
TI.~\cite{Chen2010}
Nonetheless, the reduction in the sample size inevitably induces a gap in the metallic
dispersion of TI due to the overlap of the surface states at the two opposite 
surfaces~\cite{Zhang2010,Wang2010},
which requires a detailed study of the surface or edge states.
In 2D TI, some existing works have been performed on the properties of edge
states~\cite{Konig2008,Wang2009}.
However, general understanding of the spatial features of the edge state is still
lacking.
Particularly,
we need systematic researches on the dependence of the spatial
features of the edge states on various physical parameters, which is a main
motivation of this work.

In this Letter,
we investigate theoretically the spatial variation
of the edge states in the 2D TI.
Many theoretical models have been proposed for the understanding of TI,
with the increasing demands on its research.
Kane-Mele (KM) model \cite{Kane2005,Kane2005a} is one of the prototype model of 2D TI.
This model shows a bulk energy gap on the honeycomb lattice due to the SOC, but 
its edge states on the boundaries show linear metallic dispersions 
inside of the bulk gap.
Although the established TIs are not constructed on the honeycomb lattice,
the essential physics captured in this model
is believed to
give useful insights on this field.
In this work, we focus on  the spatial dependence of the edge-state wave functions 
at the zigzag boundary. 
We define a edge-state width,
which is a convenient measure of how tightly the edge
state is confined at the boundary.
It is found that the edge states merge into the energy band of the bulk states
through a bifurcation(BFC) behavior.
The edge-state width is computed for various SOCs and sublattice potentials.
We also discuss what significant effects the variation of the edge-state
width to the boundary has in experiments.

\begin{figure}[tb]
	\includegraphics[width=8cm]{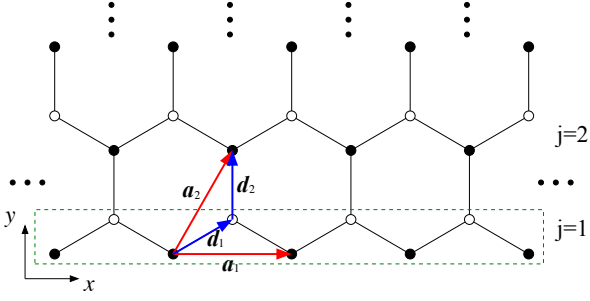}
	\caption{ \label{fig:edge} Schematic diagram for the semi-infinite
	honeycomb lattice with the boundary at the bottom.  
The lattice are described by the 
two primitive vectors, 
$\bm{a}_1{=}a(1,0)$, and
$\bm{a}_2{=}\frac{a}{2}(1, \sqrt{3})$
with two bases $(0,0)$ and $\frac{a}{2}(1,1/\sqrt{3})$
in a unit cell.}
\end{figure}

We start from the Hamiltonian of the KM model \cite{Kane2005} with
creation(annihilation) operators, 
$c_{i\sigma}^{}(c_{i\sigma}^\dagger)$ of an electron of spin
$\sigma$ at site $i$ on the honeycomb lattice.  
\begin{equation}
H=-t\!\!\sum_{\langle i,j\rangle\sigma}\!\! c_{i\sigma}^\dagger c_{j\sigma}^{}
+\lambda_v\!\! \sum_{i\sigma}\mu_i c_{i\sigma}^{\dag}c_{i\sigma}
+i\lambda_{SO}\!\!\!\!\!\!\!\!\sum_{\langle\langle i,j\rangle\rangle\alpha,\beta}
\!\!\!\!\!\!\nu_{ij}\sigma_{\alpha\beta}^{z}c_{i\alpha}^\dagger c_{j\beta}^{},
\label{KM-model}
\end{equation}
The first term describes the hopping of electrons between the nearest-neighbor
sites $\langle i,j \rangle$ with the hopping integral $t$.
The staggered sublattice potential of strength $\lambda_v$ is included in 
the second term with $\mu_i{=}\pm 1$ on each sublattice, respectively.
The third term is a spin-dependent hopping between next-nearest neighbors
$\langle\langle i,j \rangle\rangle$ due to the SOC.
$\lambda_{SO}$ is SOC,
and $\sigma^z$ is the $z$ component of Pauli matrix.
$\nu_{ij}=\pm 1$ is determined by 
$ \nu_{ij} = \frac{2}{\sqrt{3} a^2}({\bm d}_1\times{\bm d}_2)$,
where $\bm{d}_1$ and $\bm{d}_2$ are the two
vectors 
connecting next nearest neighbors $i$ and $j$ from the site $j$ to $i$.
Henceforth, we set $a{=}1$ for simplicity.
The Hamiltonian in Eq.~\eqn{KM-model}
gives the energy spectrum.
\begin{equation}
E_\sigma (\bm{k}) = \pm \sqrt{\varepsilon_0({\bm k}) +
\left\{2\sigma\lambda_{SO} 
\varepsilon_1({\bm k})+\lambda_v\right\}^2}
\end{equation}
where 
$\varepsilon_0({\bm k}) 
= 3+2\cos k_x + 4\cos\frac{k_x}{2}\cos\frac{\sqrt{3}}{2}k_y$
and
$\varepsilon_1({\bm k}) = \sin k_x-2\sin\frac{k_x}{2}\cos\frac{\sqrt{3}}{2}k_y $,
with $\sigma = \pm 1$ depending on the electron spin.
It is known that
the system has
finite gaps of $2|3\sqrt{3}\lambda_{SO}\pm\lambda_v|$ 
at two Dirac points ${\bf K}$ and $\bf K'$,
indicating that the system lies in the insulating state.

One prominent characteristic of the topological nontrivial phase
is that a metallic state emerges 
with being localized at the boundary.
In order to investigate the edge state, we assume a zigzag boundary
along the $x$ direction while the system is semi-infinite in the $y$ direction
as in \fig{fig:edge}.
It is convenient to write the edge-state wave function 
in the following form~\cite{Konig2008,Wang2009},
\begin{equation}
\Psi_{k\sigma,y} = \left(e^{ik/2}\Lambda\right)^j \Psi_{k\sigma,y=0},
\label{eq:psikj}
\end{equation}
where $\Psi_{k\sigma,y}$ is the two-component vector whose elements represent
the wave functions of spin $\sigma$
for the two bases in the unit cell. 
Here $k$ is the momentum in the $x$-direction, and $y\equiv j (\sqrt{3} a/2)$
with $j{=}1,2,\ldots$ is the site index
in the $y$-direction.
Here, $\Lambda$ is a complex number
the magnitude of which should be less than unity for
the states localized at the bottom
boundary to make the wave function decays rapidly as $y$ increases.
The effective Hamiltonian for the edge state in Eq.~\eqn{eq:psikj} is then given
by
\begin{widetext}
\begin{equation}
\label{eq:hamil}
\hat{H}_{\mbox{\tiny edge}}
=
\left[
\begin{matrix}
4 \lambda_{SO}  \sin\frac{k}{2}
\left\{ \cos\frac{k}{2}-\frac{\Lambda +\Lambda^{-1}}{2}\right\}
+ \lambda_v
&
-t e^{-ik/2}\left(2\cos\frac{k}{2}
+\Lambda^{-1}\right)
\\
-t e^{ik/2}\left(2\cos\frac{k}{2}
+\Lambda^{}\right)
&
-4 \lambda_{SO}  \sin\frac{k}{2}
\left\{ \cos\frac{k}{2}-\frac{\Lambda +\Lambda^{-1}}{2}\right\}
- \lambda_v
\end{matrix}
\right]
\end{equation}
\end{widetext}
for spin-up.
The Hamiltonian for spin-down can be obtained by substituting
$-\lambda_{SO}$ for $\lambda_{SO}$.
\begin{figure}[tbh]
	\includegraphics[width=8cm]{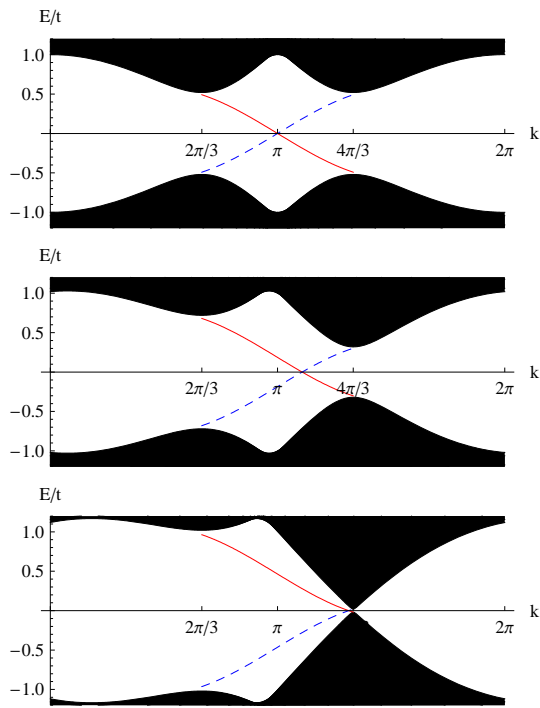}
	\caption{
	\label{fig:dispersion} 
(color online)
Edge-state energy dispersions for spin-up electrons with $\lambda_{SO}/t{=}0.1$ and
(top) $\lambda_v/t {=} 0.0$; (middle) $0.2$; (bottom) $0.49$.    
The edge-state dispersions at the lower boundary calculated in the text 
are denoted by (red) solid lines. 
Those at the upper boundary are also plotted by (blue) dotted lines for
completeness.
The shaded area
represents the bulk energy spectrum. 
}
\end{figure}
The following equation should be satisfied for
the energy, $E$.
\begin{equation}
	\label{eq:eig}
	\left|\hat{H}_{\mbox{\tiny edge}}(\Lambda)-E\right|=0
\end{equation}
It turns out that the eigenvalue equation in Eq.~\eqn{eq:eig}
is a quadratic equation of $\Lambda+\Lambda^{-1}$, yielding four solutions for
$\Lambda$, i.e., $\Lambda_1$, $\Lambda_1^{-1}$, $\Lambda_2$,
and $\Lambda_2^{-1}$, with given $k$ and $E$.
We can assume that $|\Lambda_2|\le|\Lambda_1|  \le 1$ with the loss of generality.
Only the two values $\Lambda_1$ and $\Lambda_2$ are relevant for the description of
states localized at the edge of our interest.
It is clear that the other two, $\Lambda_1^{-1}$ and $\Lambda_2^{-1}$, 
correspond to localized states at the upper boundary, if any.
The wave function
can be written as
\begin{equation}
\Psi_{k,y} = C_1 \left(e^{ik/2}\Lambda_1\right)^j\Phi_{k 1}
+ C_2 \left(e^{-ik/2}\Lambda_2\right)^j\Phi_{k 2}
\end{equation}
where $\Phi_{k i}$ is the corresponding eigenvector of the 
solution $\Lambda_i$ and the energy $E$ in \eqn{eq:eig}.
The wave function vanishes at the boundary of the sample $(j{=}0)$, 
which implies that the two eigenvectors are linearly dependent.
The condition for the linear dependence of eigenvectors produces an 
additional equation
\begin{equation}
\label{eq:BC}
E-\lambda_v+4\lambda_{SO}\sin\frac{k}{2}\left(\frac{\Lambda_{1}+\Lambda_{2}}{2}
+\left(\Lambda_{1}\Lambda_{2}-2\right)\cos\frac{k}{2}\right)=0.
\end{equation}

Simple numerical solutions of the coupled equations 
from \eqn{eq:eig} and \eqn{eq:BC}
produce the energy dispersions of the edge state as well as the corresponding
wave functions. 
In Fig.\ref{fig:dispersion} we show the edge-state energy 
dispersion for spin-up electrons with various $\lambda_{v}$.
The TRS in the system guarantees that
the dispersions for the opposite spin are obtained by the inversion of $k$ 
around $k{=}\pi$(not shown).
The resulting dispersions clearly demonstrate that
a metallic edge state exists inside the finite band gap
and 
that its dispersion relation gradually merges into the bulk states
as its momentum approaches 
$k{=}\frac{2\pi}{3}$ or $\frac{4\pi}{3}$.
Even when
the presence of $\lambda_v$ induces a difference between the gaps at the two Dirac
points and
the dispersion becomes asymmetric around $k{=}\pi$, 
the metallic edge state 
is still preserved.
Finally one of the gaps closes when $\lambda_v$ reaches the critical value 
$3\sqrt{3}\lambda_{SO}$.
This makes the whole system metallic.
Further increase of $\lambda_v$ above the critical value in turn opens a gap in the bulk. 
In this case, however, the bulk is a normal insulator which is accompanied by a
gap at the edge-state as well.
All the features of the edge-state dispersions are fully consistent with the
edge-state characteristics 
studied by the full numerical diagonalization of 
 the KM model in the strip geometry of finite width~\cite{Kane2005,Kane2005a}. 

\begin{figure}
	\includegraphics[width=8cm]{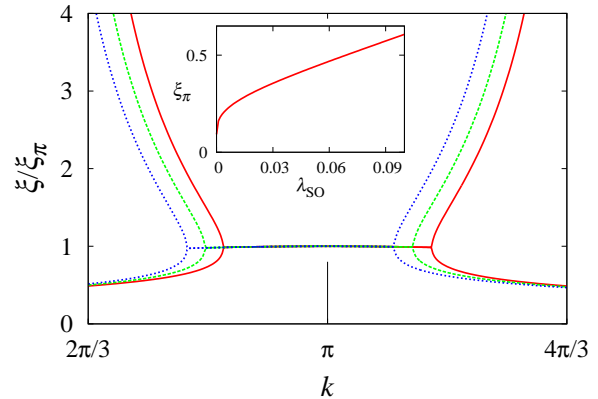}
	\caption{\label{fig:locallength}
(color online)
	Edge-state width $\xi$ as a function of momentum $k$ for
$\lambda_{SO}/t=0.1$.
The data for $\lambda_v/t=0, 0.2$ and $0.4$ are denoted by (red) solid, (green)
dashed, and (blue) dot-and-dash lines, respectively.  
The inset shows the edge-state width $\xi_\pi$ for $k{=}\pi$ 
as a function of $\lambda_{SO}$.}
\end{figure}

Our approach also allows the calculation of the wave function of the edge state, which
is useful for the study of the edge-state confinement to the boundary.
Since the system considered is semi-infinite in the transverse direction,
the result is free of the size effects due to a finite width of the system.
We define a edge-state width $\xi_i$ for each decaying state
from the $\Lambda_i$ in Eq.~\eqn{eq:psikj}.
\begin{equation}
\xi_i(k) \equiv \frac{\sqrt{3}}{2} [\ln |1/\Lambda_i(k)|]^{-1},
\end{equation}
indicating a decay $ \sim e^{-y/\xi_i(k)}$.
For $k{=}\pi$, we can solve Eqs.~\eqn{eq:eig} and \eqn{eq:BC} 
analytically,
yielding
\begin{equation}
	\xi_\pi = \frac{\sqrt{3}}{2}\left[\ln\left\{\sqrt{1+\left(\frac{t}{4\lambda_{SO}}\right)^2}
+\frac{t}{4\lambda_{SO}}\right\}\right]^{-1}.
\label{eq:length0}
\end{equation}
It is of interest to note that for $k{=}\pi$ and $|\lambda_v| < 4|\lambda_{SO}|$
the staggered lattice potential $\lambda_v$ changes only the phase factor of $\Lambda$ 
without any effect on $\xi_\pi$. (
The effect of $\lambda_v$ on $\xi$ shows up for $k{\neq}\pi$ or larger $\lambda_v$,
which will be discussed later. )
The inset of \fig{fig:locallength} shows the behavior of $\xi_\pi$ as a function of 
$\lambda_{SO}$.
The decrease in $\lambda_{SO}$ makes the wave function narrower.
Finally, the edge state is maximally confined at the boundary for $\lambda_{SO}
{=} 0$, which
corresponds to a well-known flat-band edge state of the graphene nanoribbon with a zigzag
edge.~\cite{Fujita1996JPSJ}
Our results prove that $\xi$ vanishes logarithmically $\sim
1/\ln(\lambda_{SO}/t)$ as $\lambda_{SO}$ approaches zero.
 
The edge-state width $\xi_1(k)$ and $\xi_2(k)$ are shown in \fig{fig:locallength}
for several values of $\lambda_v$. 
In the finite region around $k{=}\pi$, we find that  $\xi_1 {=} \xi_2$ 
and that the value is rather insensitive to the variation of $k$.
Above a certain momentum, $\xi_1$ is no longer identical to $\xi_2$ and
increases rapidly as $k$ is increased with the monotonic decrease of $\xi_2$.
Such behavior of the edge-state width is reminiscent of a bifurcation (BFC). 
Careful examination of Eq.~\eqn{eq:eig} gives us some insight on the origin of
the BFC:
A quadratic equation of $\Lambda+\Lambda^{-1}$ with real coefficients give two
complex roots or two real roots.
In the former case, 
two complex roots, which are complex conjugates to each other, enforce
$\Lambda_1{=}\Lambda_2^*$, which explains $\xi_1{=}\xi_2$ around $k{=}\pi$.
On the other hand, such enforcement is not imposed for the latter case and
$\xi_1{>}\xi_2$.
Accordingly,
the BFC occurs when Eq.~\eqn{eq:eig} has one real double root.
It is also worthwhile to mention that $\xi_1$ diverges at two Dirac points,
where the edge state merges to the energy band of the bulk state.
We can verify that at $k{=}\pi\pm\pi/3$ the coupled equations in \eqn{eq:eig} and
\eqn{eq:BC} are satisfied by the energy $E{=}\mp3\sqrt{3}\lambda_{SO}+\lambda_v$
and $\Lambda_1=\pm1$, 
which guarantees the divergence of $\xi_1$ and the merging of the edge state to
the energy band of extended states at two Dirac points.

\begin{figure}
	\includegraphics[width=8cm]{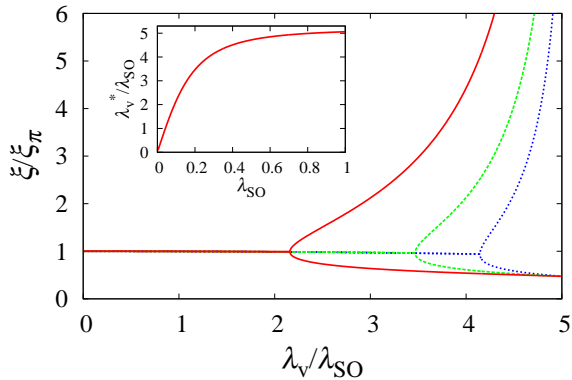}
	\caption{\label{fig:Fermi_state} (color online)
	The edge-state width $\xi$ at $E=0$
	as a function of the ratio of the sublattice potential $\lambda_{v}$  to
SOC $\lambda_{SO}$. 
The data for $\lambda_{SO}/t=0.1, 0.2,$ and $0.3$ are denoted by (red) solid,
(green) dashed, and (blue) dot-and-dash lines, respectively.
	The inset shows the location of the BFC value $\lambda_v^\ast$
in units of $\lambda_{SO}$  as a function of $\lambda_{SO}$.}
\end{figure}

The edge-state width has great significance in
the experiments on the samples of a finite width. 
When $\xi$ is comparable to the sample width,
the overlap of the edge states on the opposite boundaries becomes
significant and develops a finite gap in the edge-state dispersions.
Such effects have already been investigated by the variation of the film
thickness~\cite{Zhang2010,Wang2010}.

In experiments
it would be more convenient to tune the edge-state width  by the variation 
of external fields in the sample of fixed width.
Silicene, which was recently
synthesized~\cite{Lalmi2010apl,Vogt2012prl,Lin2012ape},
provides one interesting possibility.
The buckling of two sublattices in silicene enables one to control
the sublattice potential $\lambda_v$ by applying external electric
fields~\cite{Ezawa2012prl}.
The variation of the edge-state width by the change of $\lambda_v$, shown
in \fig{fig:Fermi_state}, exhibits a remarkable dependence, particularly around
a BFC point. 
Such a dependence is expected to give observable effects on the edge-state gap
in narrow samples by the application of external electric fields.
It is also interesting that 
the BFC point
$\lambda_v^\ast$ where $\xi(\lambda_v)$ bifurcates
is enhanced by the increase of SOC,
which implies that 
strong SOC pins down the width $\xi$ and suppresses a formation of an
edge-state gap up to a higher BFC point $\lambda_v^\ast$. 

\begin{figure}
	\includegraphics[width=8cm]{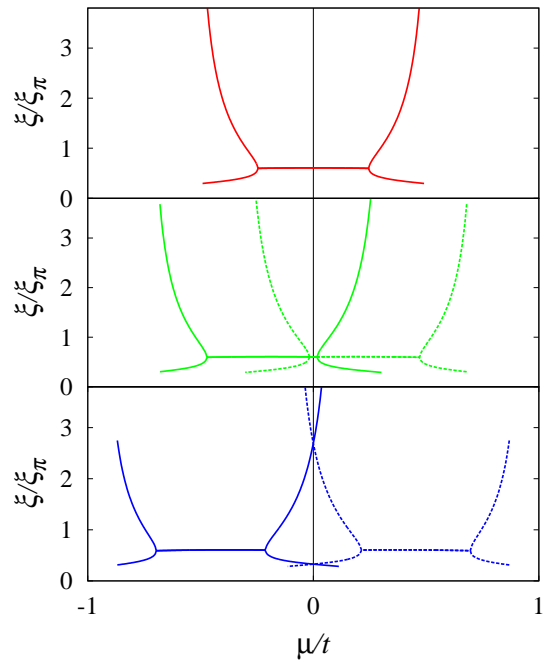}
	\caption{\label{fig:length_mu} 
The width of the edge-state for spin-up electrons as
a function of the chemical potential $\mu$ for (top) $\lambda_v/t=0$; (middle) 0.2; (bottom)
0.4. 
The solid line corresponds to the edge states at the lower boundary and the
dotted line to those at the upper boundary.}
\end{figure}

An alternative way to tune the edge-state width in experiments 
is to control the chemical potential by the change of a gate voltage~\cite{Chen2010}.
In such experiments the dependence of $\xi$  on the chemical
potential shown in \fig{fig:length_mu} is useful.
We have assumed the band structures are not changed significantly by the variation
of chemical potential. 
In the absence of the staggered sublattice potential,
$\xi$ is hardly changed
by a small variation of the chemical potential around the half-filling case ($\mu{=}0$). 
Only when the chemical potential change exceeds,
a BFC point $\xi$ increases rapidly and a gap formation is allowed
with a proper momentum transfering perturbation.
A small increase of $\lambda_v$ shifts BFC points for both boundaries
closer to the Fermi level of a half-filled system, which
makes it easier to delocalize the edges states by the gate voltage.
For large $\lambda_v$,
the edge states at the both boundaries get wider even in the
half-filling case. 
It is interesting that
the change in the chemical potential increases the width of one
edge state while that of the other one is suppressed.


In summary,
we have derived analytically the equations for the energies 
and the width of the edge states
in the KM model with a zigzag edge.
Through the analysis of the width of the edge states,
we have shown that
a BFC of the edge-state width plays an important role 
in the properties of the edge states.
It has also been shown that the effects of such a BFC can be
uncovered by the adjustment of experimentally controllable parameters such as the sublattice potential or the chemical potential.
\begin{acknowledgments}
This work was supported by the National Research Foundation of Korea (NRF)
funded by the Korean government (MEST) through the Quantum Metamaterials
Research Center, Grant No. 2011-0000982 (G.S.J.), and Basic Science Research,
Grant No. 2010-0010937 (G.S.J) and Grant No. 2011-0018306 (H.D.). 
\end{acknowledgments}

\bibliographystyle{apsrev}
\bibliography{TI_Graphene}

\end{document}